\documentclass[aps,twocolumn,superscriptaddress]{revtex4}
\usepackage{times}
\usepackage{color}
\usepackage{float}
\usepackage{amssymb}
\usepackage{epsfig}
\usepackage{booktabs}
\usepackage{mathrsfs}
\usepackage{amsmath}

\begin{document}

\title{Quantum corrections to the entropy and its application in the study of quantum Carnot engines}

\author{Tian Qiu}
\affiliation{School of Physics, Peking University, Beijing, 100871, China}

\author{Zhaoyu Fei}
\affiliation{School of Physics, Peking University, Beijing, 100871, China}

\author{Rui Pan}
\affiliation{School of Physics, Peking University, Beijing, 100871, China}

\author{Haitao Quan}
\affiliation{School of Physics, Peking University, Beijing, 100871, China}
\affiliation{Collaborative Innovation Center of Quantum Matter, Beijing 100871, China}
\affiliation{Frontiers Science Center for Nano-optoelectronics, Peking University, Beijing, 100871, China}

\date{\today}

\begin{abstract}
 Entropy is one of the most basic concepts in thermodynamics and statistical mechanics. The most widely used definition of statistical mechanical entropy for a quantum system is introduced by von Neumann. While in classical systems, the statistical mechanical entropy is defined by Gibbs. The relation between these two definitions of entropy is still not fully explored. In this work, we study this problem by employing the phase-space formulation of quantum mechanics. For those quantum states having well-defined classical counterparts, we study the quantum-classical correspondence and quantum corrections of the entropy. We expand the von Neumann entropy in powers of ${\hbar}$ by using the phase-space formulation, and the zeroth order term reproduces the Gibbs entropy. We also obtain the explicit expression of the quantum corrections of the entropy. Moreover, we find that for the thermodynamic equilibrium state, all terms odd in ${\hbar}$ are exactly zero. As an application, we derive quantum corrections for the net work extraction during a quantum Carnot cycle. Our results bring important insights to the understanding of quantum entropy and may have potential applications in the study of quantum heat engines.
\end{abstract}

\maketitle
\section{Introduction}
Entropy is, without any doubt, one of the most important physical concepts in thermodynamics and statistical mechanics. The notion of thermodynamic entropy was first introduced by Clausius in 1865 \cite{Clausius1865}, and Boltzmann gave the statistical interpretation of the entropy in 1877, i.e., the relation between thermodynamic entropy and probability theory \cite{Boltzmann1877a, Boltzmann1877b}. In classical statistical mechanics, the macroscopic state of a system is characterized by a distribution of the microstates, and the statistical mechanical entropy of this distribution is first introduced by Gibbs in 1902 \cite{Gibbs1902, Jaynes1965, Callen1985}. Relevantly, a quantum statistical mechanical entropy for quantum systems was proposed by von Neumann in 1927 \cite{von Neumann1927, von Neumann1955, Fano1957}. In the literature, there are some discussions on the properties of both classical and quantum entropy \cite{Wehrl1978, Wehrl1979, Zachos2007}. For example, in Ref. \cite{Wehrl1979}, a new definition of classical entropy (Wehrl entropy) is proposed and its relation to quantum entropy is discussed. In Ref. \cite{Zachos2007}, a classical bound on the quantum entropy is proposed. However, how the definitions of entropy in these two regimes are related to each other is not fully explored.

The phase-space formulation of quantum mechanics is equivalent to Hilbert space formulation and Feynman's path integral formulation of quantum mechanics \cite{Wigner1932, Hillery1984, Polkovnikov2010}. Using the Weyl-Wigner transform, we map operators in the Hilbert space formulation into Weyl symbols (functions) in the phase-space formulation. Then the evolution equation of operators, as well as their expectation values, can be reformulated in the phase space. In Ref. \cite{Wigner1932, Nienhuis1970, OConnell1985}, the quantum corrections for the thermodynamic equilibrium states were studied by using the phase-space formulation. It turns out that if we expand the Wigner function of a quantum thermodynamic equilibrium state in powers of ${\hbar}$, the zeroth order term reproduces the thermodynamic equilibrium distribution of its classical counterpart, and all terms odd in ${\hbar}$ are exactly zero. In Ref. \cite{Beretta1984, Wang1986}, the quantum-classical correspondence of entropy has been studied. Recently, the quantum corrections of work statistics in closed systems have been derived in Ref. \cite{Fei2018}. Nevertheless, the quantum corrections to the entropy has not been explored systematically, probably due to the difficulty in the Weyl-Wigner transform of the logarithm of the density matrix.

In this article, in parallel to the work of Wigner in 1932 \cite{Wigner1932}, we study the quantum corrections of the entropy by utilizing the phase-space formulation of quantum mechanics. For those quantum states having well-defined classical counterparts, we expand the von Neumann entropy in powers of ${\hbar}$ in the phase space representation, and obtain the quantum corrections to the entropy. Specially, for the thermal equilibrium states, we prove that all terms odd in ${\hbar}$ are exactly zero. As an application, we derive quantum corrections of the net work extraction during a quantum Carnot cycle. Our results bring important insights to the understanding of quantum entropy.

This article is organized as follows. In Sec. II, we derive the quantum corrections to the entropy of states which have well defined classical counterparts. We prove that if we expand the von Neumann entropy in powers of ${\hbar}$, the zeroth order term reproduces the classical Gibbs entropy, and we obtain the analytical expressions for the first and the second order quantum corrections. In Sec. III, we take the thermal equilibrium state as an example, and we find that all correction terms odd in ${\hbar}$ are exactly zero. As an application, we study the quantum correction of the net work extraction during a quantum Carnot cycle in Sec. IV. We conclude our paper in Sec. V.

\section{Quantum corrections to the entropy for states having classical counterparts}
Let us consider a density matrix ${\hat{\rho}}$. We are interested in calculating the quantum correction to the classical Gibbs entropy. As is known, the most widely used definition of the entropy is given by von Neumann \cite{Wehrl1978}. Hereafter, we use von Neumann entropy and quantum entropy synonymously. The quantum entropy of the density matrix is given by \cite{von Neumann1927, von Neumann1955}
\begin{equation}\label{eq:004}
S_q=-\rm{Tr}\left[\hat{\rho}\ln\hat{\rho}\right].
\end{equation}
Here, we have set the Boltzmann's constant to be ${1}$, thus the entropy becomes dimensionless.

In order to calculate the quantum correction to the entropy, an intuitive idea is to reformulate the above expression (\ref{eq:004}) with the phase-space formulation of quantum mechanics, and expand it in powers of ${\hbar}$. However, it turns out that this approach does not work due to the fact that the Weyl symbol of ${\ln\hat{\rho}}$ is not equal to the logarithm of the Weyl symbol of ${\hat{\rho}}$ \cite{Zachos2007, OConnell2006}. Hence, in order to calculate the quantum correction to the entropy, we have to expand ${\ln\hat{\rho}}$ in powers of ${\hbar}$ through other alternative methods. In the following, we will introduce our method.

Let us consider a function of the operator ${\hat{\rho}}$
\begin{equation}\label{eq:001}
\hat{f}(a,b)=\ln(a\hat{\rho}+b),
\end{equation}
where ${a}$ and ${b}$ are two constants. Obviously ${\hat{f}(a,b)}$ is the unique solution to the following equations about ${a}$,
\begin{equation}\label{eq:002}
\left\{
\begin{aligned}
&\frac{\partial \hat{f}(a,b)}{\partial a}=\hat{A} \\
&\hat{f}(0,b)=\ln b,
\end{aligned}
\right.
\end{equation}
where
\begin{equation}\label{eq:003}
\hat{A}=\frac{\hat{\rho}}{a\hat{\rho}+b}
\end{equation}
is another operator. Accordingly, the von Neumann entropy of ${\hat{\rho}}$ can be expressed as a function of ${\hat{f}(a,b)}$ as
\begin{equation}\label{eq:005}
S_q=-\!\lim_{a\!\to\!1,b\!\to\!0}\langle \hat{f}(a,b)\rangle.
\end{equation}
Here the angular bracket depicts the ensemble average with respect to ${\hat{\rho}}$. Eq. (\ref{eq:002}) and Eq. (\ref{eq:005}) can be rewritten as follows by using the phase-space formulation \cite{Hillery1984}
\begin{equation}\label{eq:006}
\left\{
\begin{aligned}
&\frac{\partial f_w(a,b;x,p)}{\partial a}=A_w(a,b;x,p) \\
&f_w(0,b;x,p)=\ln b
\end{aligned}\ \ ,
\right.
\end{equation}
\begin{equation}\label{eq:007}
S_q=-\!\lim_{a\!\to\!1,b\!\to\!0}\int\frac{dxdp}{2\pi\hbar}W(x,p)f_w(a,b;x,p).
\end{equation}
Here, the subscript ``${w}$" indicates the Weyl symbols of these operators, and ${W(x,p)}$ is the Weyl symbol of the density matrix ${\hat{\rho}}$ (Wigner function) of the system, which is defined as \cite{Wigner1932}
\begin{equation}\label{eq:008}
W(x,p):=\int dy \left\langle x-\frac{y}{2}\Big| \hat{\rho}\Big| x+\frac{y}{2}\right\rangle e^{\frac{ipy}{\hbar}}.
\end{equation}
Also, according to Eq. (\ref{eq:003}), we can find that ${A_w}$ satisfies the following equation,
\begin{equation}\label{eq:009}
(aW+b)\star A_w=W.
\end{equation}
Here ``${\star}$" is the Moyal product \cite{Groenewold1946} which is defined as
\begin{equation}\label{eq:010}
\star:=e^{-\frac{i\hbar}{2}(\stackrel{\leftarrow}{\partial_p}\stackrel{\rightarrow}{\partial_x}
-\stackrel{\leftarrow}{\partial_x}\stackrel{\rightarrow}{\partial_p})},
\end{equation}
where the arrows indicate the directions the derivatives act upon.

\begin{table*}
	\centering
	\caption{Partial differential equations of ${f^{(0)}_w}$, ${f^{(1)}_w}$, ${f^{(2)}_w}$ and their solutions.}
\renewcommand\tabcolsep{19.0pt}
	\begin{tabular}[c]{c||c}
		\hline
        \hline
\begin{minipage}{2cm}\vspace{0.3cm}\vspace{0.3cm}\end{minipage}
	Equations for ${f^{(0)}_w}$, ${f^{(1)}_w}$, ${f^{(2)}_w}$ & Solutions for ${f^{(0)}_w}$, ${f^{(1)}_w}$, ${f^{(2)}_w}$ \\
		\hline
        \hline
\begin{minipage}{2cm}\vspace{0.7cm}\vspace{0.7cm}\end{minipage}
        ${\begin{aligned}
		&\frac{\partial f^{(0)}_w(a,b)}{\partial a}=A^{(0)}_w \\
        &f^{(0)}_w(0,b)=\ln b
        \end{aligned}}$ &
        ${\begin{aligned}
        &f^{(0)}_w(a,b)=\ln(aW^{(0)}+b)
        \end{aligned}}$ \\
		\hline
\begin{minipage}{2cm}\vspace{0.7cm}\vspace{0.7cm}\end{minipage}
        ${\begin{aligned}
        &\frac{\partial f^{(1)}_w(a,b)}{\partial a}=A^{(1)}_w \\
        &f^{(1)}_w(0,b)=0
        \end{aligned}}$ &
        ${\begin{aligned}
        f^{(1)}_w=\frac{aW^{(1)}}{aW^{(0)}+b}
        \end{aligned}}$ \\
        \hline
\begin{minipage}{2cm}\vspace{0.7cm}\vspace{0.7cm}\end{minipage}
        ${\begin{aligned}
        &\frac{\partial f^{(2)}_w(a,b)}{\partial a}=A^{(2)}_w \\
        &f^{(2)}_w(0,b)=0
        \end{aligned}}$ &
        ${\begin{aligned}
        f^{(2)}_w=\frac{aW^{(2)}}{aW^{(0)}+b}-\frac{a^2[8(W^{(1)})^2+W^{(0)}(\stackrel{\leftarrow}{\partial_p}\stackrel{\rightarrow}{\partial_x}
-\stackrel{\leftarrow}{\partial_x}\stackrel{\rightarrow}{\partial_p})^2W^{(0)}]}{16(aW^{(0)}+b)^2}+\frac{a^3G(W^{(0)})}{12(aW^{(0)}+b)^3}
        \end{aligned}}$ \\
        \hline
        \hline
	\end{tabular}
\end{table*}

For a quantum state which has a classical counterpart \cite{footnote1}, the Wigner function can be expanded in nonnegative powers of ${\hbar}$ as
\begin{equation}\label{eq:011}
W=W^{(0)}+i\hbar W^{(1)}+(i\hbar)^2 W^{(2)}+o(\hbar^2),
\end{equation}
where ${W^{(0)}(x,p)}$ is the corresponding classical probability distribution in the phase space. From Eq. (\ref{eq:009}) and Eq. (\ref{eq:011}), we find that there is no term in negative powers of ${\hbar}$ in ${A_w}$ \cite{footnote2}. Then ${A_w}$ can be expanded in powers of ${\hbar}$ as
\begin{equation}\label{eq:012}
A_w=A^{(0)}_w+i\hbar A^{(1)}_w+(i\hbar)^2 A^{(2)}_w+o(\hbar^2).
\end{equation}
Also, the Moyal product can be expanded in powers of ${\hbar}$ \cite{Fei2018}. Substituting Eq. (\ref{eq:011}) and Eq. (\ref{eq:012}) into Eq. (\ref{eq:009}), after some algebra, one can obtain the explicit expressions for ${A^{(0)}_w}$, ${A^{(1)}_w}$ and ${A^{(2)}_w}$,
\begin{equation}\label{eq:013}
A^{(0)}_w=\frac{W^{(0)}}{aW^{(0)}+b},
\end{equation}
\begin{equation}\label{eq:014}
A^{(1)}_w=\frac{bW^{(1)}}{(aW^{(0)}+b)^2},
\end{equation}
\begin{widetext}
\begin{equation}\label{eq:015}
A^{(2)}_w=\frac{bW^{(2)}}{(aW^{(0)}+b)^2}-\frac{ab(W^{(0)})^2}{(aW^{(0)}+b)^3}-\frac{a}{8}\left[\frac{b(aW^{(0)}+b)[W^{(0)}(\stackrel{\leftarrow}{\partial_p}\stackrel{\rightarrow}{\partial_x}
-\stackrel{\leftarrow}{\partial_x}\stackrel{\rightarrow}{\partial_p})^2W^{(0)}]-2abG(W^{(0)})}{(aW^{(0)}+b)^4}\right],
\end{equation}
where
\begin{equation}\label{eq:017}
G(W^{(0)})=\partial^2_p W^{(0)}(\partial_x W^{(0)})^2+\partial^2_x W^{(0)}(\partial_p W^{(0)})^2
-2\partial_x W^{(0)}\partial_p W^{(0)}\partial_{xp} W^{(0)}.
\end{equation}
\end{widetext}

Here ${A^{(0)}_w}$ is the classical counterpart of ${\hat{A}}$, and ${A^{(1)}_w}$, ${A^{(2)}_w}$ are the quantum corrections to the classical counterpart.

We also expand ${f_w}$ in powers of ${\hbar}$ as \cite{footnote3}
\begin{equation}\label{eq:018}
f_w=f^{(0)}_w+i\hbar f^{(1)}_w+(i\hbar)^2 f^{(2)}_w+o(\hbar^2).
\end{equation}
One can obtain ${f^{(0)}_w}$, ${f^{(1)}_w}$ and ${f^{(2)}_w}$ from solving Eq. (\ref{eq:006}), and the solutions are listed in Table I. From Eq. (\ref{eq:007}), Eq. ({\ref{eq:011}), and Eq. (\ref{eq:018}), one can find that there is no term in negative powers of ${\hbar}$ in ${S_q}$. Then we expand the von Neumann entropy in powers of ${\hbar}$ as
\begin{equation}\label{eq:019}
S_q=S^{(0)}_q+i\hbar S^{(1)}_q+(i\hbar)^2 S^{(2)}_q+o(\hbar^2).
\end{equation}
Substituting the results in Table I and Eq. (\ref{eq:011}) into Eq. (\ref{eq:007}), we obtain
\begin{equation}\label{eq:020}
S^{(0)}_q=-\int\frac{dxdp}{2\pi\hbar}\ W^{(0)} \ln W^{(0)}.
\end{equation}
One can see that ${S^{(0)}_q}$ is the classical Gibbs entropy \cite{Gibbs1902}, i.e., the von Neumann entropy reproduces the classical Gibbs entropy when we take the limit ${\hbar\to 0}$ \cite{Beretta1984, Wang1986}. In addition, from Eq. (\ref{eq:019}), one can obtain the quantum corrections to the classical Gibbs entropy. ${S^{(1)}_q}$ and ${S^{(2)}_q}$ are the first and the second order quantum corrections of entropy respectively. Their expressions can be given explicitly as follows
\begin{equation}\label{eq:021}
S^{(1)}_q=-\int\frac{dxdp}{2\pi\hbar}\ \left[W^{(1)} \ln W^{(0)}+W^{(1)}\right],
\end{equation}
\begin{widetext}
\begin{equation}\label{eq:022}
S^{(2)}_q=-\int\frac{dxdp}{2\pi\hbar}\ \left[W^{(2)} \ln W^{(0)}+W^{(2)}
+\frac{(W^{(1)})^2}{2W^{(0)}}-\frac{W^{(0)}(\stackrel{\leftarrow}{\partial_p}\stackrel{\rightarrow}{\partial_x}
-\stackrel{\leftarrow}{\partial_x}\stackrel{\rightarrow}{\partial_p})^2W^{(0)}}{16W^{(0)}}+\frac{G(W^{(0)})}{12(W^{(0)})^2}\right].
\end{equation}
\end{widetext}
We would like to emphasize that Eqs. (\ref{eq:020}-\ref{eq:022}) are the main results of our paper, i.e., the quantum corrections to the entropy.
\begin{table*}
	\centering
	\caption{Expressions of ${W^{(0)}_{eq}}$, ${W^{(1)}_{eq}}$, ${W^{(2)}_{eq}}$ and expressions of ${S^{(0)}_q}$, ${S^{(1)}_q}$, ${S^{(2)}_q}$ from Eqs. (\ref{eq:020}-\ref{eq:022})}
\renewcommand\tabcolsep{14.7pt}
	\begin{tabular}[c]{c||c}
		\hline
        \hline
\begin{minipage}{2cm}\vspace{0.3cm}\vspace{0.3cm}\end{minipage}
	Expressions for ${W^{(0)}_{eq}}$, ${W^{(1)}_{eq}}$, ${W^{(2)}_{eq}}$ \cite{Wigner1932}& Expressions for ${S^{(0)}_q}$, ${S^{(1)}_q}$, ${S^{(2)}_q}$ from Eq. (\ref{eq:020}-\ref{eq:022}) \\
		\hline
        \hline
\begin{minipage}{2cm}\vspace{0.7cm}\vspace{0.7cm}\end{minipage}
        ${\begin{aligned}
		&W^{(0)}_{eq}(x,p)=\frac{1}{Z_{cl}} e^{-\beta\epsilon}
        \end{aligned}}$ &
        ${\begin{aligned}
        &S^{(0)}_q=S_{cl}
        \end{aligned}}$ \\
		\hline
\begin{minipage}{2cm}\vspace{0.7cm}\vspace{0.7cm}\end{minipage}
        ${\begin{aligned}
        &W^{(1)}_{eq}(x,p)=0
        \end{aligned}}$ &
        ${\begin{aligned}
        &S^{(1)}_q=0
        \end{aligned}}$ \\
        \hline
\begin{minipage}{2cm}\vspace{0.7cm}\vspace{0.7cm}\end{minipage}
        ${\begin{aligned}
        W^{(2)}_{eq}(x,p)=\frac{1}{Z_{cl}}e^{-\beta\epsilon}\Big[\eta(\beta,x,p)-\big\langle \eta(\beta,x,p)\big\rangle_{eq}\Big]
        \end{aligned}}$ &
        ${\begin{aligned}
        S^{(2)}_q=\Big(1-\beta\big\langle\epsilon(x,p)\big\rangle_{eq}\Big)\big\langle \eta(\beta,x,p)\big\rangle_{eq}
+\beta\big\langle\epsilon(x,p) \eta(\beta,x,p)\big\rangle_{eq}
        \end{aligned}}$ \\
        \hline
        \hline
	\end{tabular}
\end{table*}
\section{quantum corrections to the entropy for a thermal equilibrium state}
The above results about quantum corrections to the entropy (\ref{eq:020}-\ref{eq:022}) are valid as long as the density matrix has a well-defined classical counterpart (\ref{eq:011}). However, for some special cases, e.g., the thermal equilibrium state, the quantum corrections to the classical Gibbs entropy can be obtained through a simpler method. In the following, we will use this method to calculate the quantum corrections to the entropy for the thermal equilibrium state.

For later convenience, we introduce the system and some notations before the discussions about the entropy. For simplicity, we consider a quantum system whose Hamiltonian can be written as
\begin{equation}\label{eq:025}
\hat{H}=\frac{\hat{p}^2}{2m}+U(\hat{x}),
\end{equation}
where ${m}$ is the mass of the particle and ${U(\hat{x})}$ is the external potential. The Weyl symbol of Eq. (\ref{eq:025}) is the corresponding classical Hamiltonian,
\begin{equation}\label{eq:026}
\epsilon(x,p)=\frac{p^2}{2m}+U(x).
\end{equation}
In Ref. \cite{Wigner1932}, Wigner obtained the Weyl symbol of the exponential Hamiltonian by expanding it in powers of ${\hbar}$ when investigating the quantum corrections to the classical thermodynamical quantities (such as the kinetic energy and the potential energy):
\begin{equation}\label{eq:023}
[e^{-\beta \hat{H}}]_w=e^{-\beta \epsilon}\left[1+(i\hbar)^2 \eta(\beta,x,p)+o(\hbar^2)\right],
\end{equation}
where
\begin{equation}\label{eq:024}
\eta(\beta,x,p)=\frac{\beta^2}{8m}\left[\partial^2_x U-
\frac{\beta}{3}(\partial_x U)^2-\frac{\beta}{3m}p^2 \partial^2_x U\right].
\end{equation}

As we know, for the thermodynamic equilibrium state, one can calculate the quantum entropy indirectly from the relation between the entropy, the internal energy, and the free energy,
\begin{equation}\label{eq:030}
S_{cl}=\beta\big\langle\epsilon(x,p)\big\rangle_{eq}+\ln Z_{cl},
\end{equation}
where ${\langle...\rangle_{eq}}$ depicts an ensemble average with respect to the classical canonical distribution, and
\begin{equation}\label{eq:029}
Z_{cl}=\int\frac{dxdp}{\hbar}\ e^{-\beta\epsilon(x,p)}
\end{equation}
is the classical partition function.

Similarly, for quantum systems, the von Neumann entropy for a quantum thermal equilibrium state can be written as
\begin{equation}\label{eq:033}
S_q=\langle\beta\hat{H}\rangle+\ln Z_q,
\end{equation}
where ${\langle...\rangle}$ is the same as that in Eq. (\ref{eq:005}), and
\begin{equation}\label{eq:032}
Z_q=\rm{Tr}[e^{-\beta\hat{H}}]
\end{equation}
is the quantum partition function. In the phase-space formulation, Eq. (\ref{eq:033}) can be rewritten as
\begin{equation}\label{eq:034}
S_q=\frac{\beta}{Z_q}\int\frac{dxdp}{2\pi\hbar}\ \epsilon(x,p)[e^{-\beta \hat{H}}]_w+\ln Z_q.
\end{equation}
Also, by expanding ${1/Z_q}$ and ${\ln Z_q}$ in powers of ${\hbar}$, we have (see Eq. (\ref{eq:023}))
\begin{equation}\label{eq:035}
\left\{
\begin{aligned}
&\frac{1}{Z_q}=\frac{1}{Z_{cl}}\Big[1-(i\hbar)^2\big\langle \eta(\beta,x,p) \big\rangle_{eq}+o(\hbar^2)\Big] \\
&\ln Z_q=\ln Z_{cl}+(i\hbar)^2 \big\langle \eta(\beta,x,p) \big\rangle_{eq}+o(\hbar^2)
\end{aligned}.
\right.
\end{equation}
Finally, substituting Eq. (\ref{eq:035}) into Eq. (\ref{eq:034}), one obtains the ${\hbar}$ expansion of the von Neumann entropy for a quantum thermodynamic equilibrium state,
\begin{widetext}
\begin{equation}\label{eq:037}
S_q=S_{cl}+(i\hbar)^2\Big[\Big(1-\beta\big\langle\epsilon(x,p)\big\rangle_{eq}\Big)\big\langle \eta(\beta,x,p)\big\rangle_{eq}
+\beta\big\langle\epsilon(x,p) \eta(\beta,x,p)\big\rangle_{eq}\Big]+o(\hbar^2).
\end{equation}
\end{widetext}

We can compare the results of quantum corrections to the entropy obtained in this section with that in last section (substituting the quantum corrections to the Wigner function ${W^{(0)}_{eq}}$, ${W^{(1)}_{eq}}$, ${W^{(2)}_{eq}}$ \cite{Wigner1932} into Eqs. (\ref{eq:020}-\ref{eq:022}), we immediately obtain the quantum corrections to the entropy for the thermal equilibrium states). The results are listed in Table II. One can see that these results are exactly the same as Eq. (\ref{eq:037}), as we anticipate. It is worth mentioning that the explicit expressions of the quantum entropy as well as the classical entropy for a special case, i.e., the harmonic oscillator, has been obtained in Ref. \cite{Wang1986}. Thus the quantum corrections to the entropy for the harmonic oscillator can be obtained straightforwardly. But our result (\ref{eq:037}) is a general result and not restricted to the harmonic oscillator.

From Eq. (\ref{eq:037}), one can see that the zeroth order term is nothing but the classical Gibbs entropy, which means that the von Neumann entropy reproduces the classical Gibbs entropy in the limit of ${\hbar \to 0}$ \cite{Beretta1984, Wang1986}. Furthermore, all terms in odd powers of ${\hbar}$ are strictly zero, which is similar to the results in Refs. \cite{Wigner1932, Fei2018}. Also, an explicit expression of the lowest order correction to the entropy (${\propto\hbar^2}$) is obtained, which vanishes when the temperature of the system is high enough (${\beta\to 0}$).

Before ending this section, we would like to emphasize that Eq. (\ref{eq:037}) is applicable to thermal equilibrium states only, but Eqs. (\ref{eq:020}-\ref{eq:022}) can be applied to any state as long as it has a classical counterpart, besides the thermodynamic equilibrium state. For example, let us consider a mixed state
\begin{equation}\label{eq:038}
\hat{\rho}=\frac{1}{2}\hat{\rho}_{eq}+\frac{1}{2}D^\dagger(\alpha)\hat{\rho}_{eq}D(\alpha),
\end{equation}
where
\begin{equation}\label{eq:038a}
D^\dagger(\alpha)=e^{\alpha a^\dagger-\alpha^* a}
\end{equation}
is the displacement operator, ${\alpha}$ is an arbitrary complex number. If we set ${x_0=-(\alpha+\alpha^*)/\sqrt{2}}$ and ${p_0=(\alpha-\alpha^*)/i\sqrt{2}}$, one can easily prove that the Wigner function of ${\hat{\rho}}$ is
\begin{equation}\label{eq:039}
W(x,p)=\frac{1}{2}W_{eq}(x,p)+\frac{1}{2}W_{eq}(x-x_0,p-p_0).
\end{equation}
Similarly one can calculate the quantum corrections to the entropy for this state through Eqs. (\ref{eq:020}-\ref{eq:022}). But Eq. (\ref{eq:037}) is not applicable in this case.

\section{quantum corrections to the net work extraction during a quantum Carnot cycle}
The concept of quantum heat engines has been proposed in 1959 \cite{Scovil1959, Geusic1967, Alicki1979}, and interests in this topic have revived in the past two decades, probably due to the developments of quantum technologies \cite{Rosnagel2016, Scully2003} and its many potential applications in chemistry and biology. It can generate useful work using quantum matter as its working substance, and exhibits many unusual and exotic properties \cite{Quan2007, Quan2005, Kieu2004, Bender2000, Geva1992, Rezek2006, Vinjanampathy2016, Alicki2018}. By exploiting quantum features of the working substance, such as quantum coherence and quantum entanglement, the quantum heat engines may operate in an advantageous way in comparison with the classical heat engines.

As can be expected, in the classical limit ($\hbar\to 0$ and/or $\beta\to 0$), the performance of the quantum heat engine converges to its classical counterpart. But in the extreme quantum limit, the performance of the quantum heat engine will deviate significantly from its classical counterpart. The correction of the work extraction by a quantum heat engine in a Carnot cycle to its classical counterpart is of great interest in many situations. However, the quantum corrections to the work extrction are rarely addressed in the literature. In the following, we will calculate the quantum corrections to the work extraction of a quantum Carnot cycle by utilizing the results that we have obtained in previous sections.

\begin{figure}[htbp]
\includegraphics[width=0.40\textwidth]{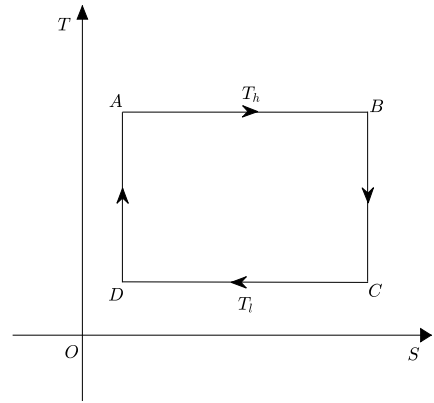}
\caption{A schematic temperature-entropy (T-S) diagram for a quantum Carnot cycle. The process from A to B (C to D) is the isothermal expansion (compression) process, in which the working substance is put in contact with the high (low) temperature ${T_h}$ (${T_l}$) heat bath. The processes from B to C and from D to A are two adiabatic processes.} \label{Figure1}
\end{figure}

Let us recall that the quantum Carnot cycle consists of two quantum isothermal processes and two quantum adiabatic processes (see Fig. 1). Without losing generality, we choose a quantum system, whose thermal equilibrium states have well-defined classical counterparts \cite{footnote4}, as our working substance, such as a particle in an external potential. We assume that the process from A to B (C to D) is the isothermal expansion (compression) process \cite{Quan2007}, in which the working substance is put in contact with the high (low) temperature ${T_h}$ (${T_l}$) heat bath, and the processes from B to C and from D to A are two quantum adiabatic processes. According to Ref. \cite{Quan2007}, the net work extraction during this quantum Carnot cycle is given by
\begin{equation}\label{eq:040}
\mathscr{W}_q=(T_h-T_l)[S_q(B)-S_q(A)].
\end{equation}
Here, ${S_q(A)}$ (${S_q(B)}$) is the von Neumann entropy of the working substance before (after) the isothermal expansion process. We consider an ideal Carnot cycle. That means the processes are quasi-static and the system is in thermal equilibrium at every moment of time. Therefore, one can obtain the quantum corrections to the net work extraction during every cycle from Eq. (\ref{eq:037}) and Eq. (\ref{eq:040}),
\begin{equation}\label{eq:041}
\mathscr{W}_q=\mathscr{W}^{(0)}_q+i\hbar \mathscr{W}^{(1)}_q+(i\hbar)^2 \mathscr{W}^{(2)}_q+o(\hbar^2),
\end{equation}
where (see Table II for the definition of ${S_q^{(2)}}$)
\begin{equation}\label{eq:042}
\mathscr{W}^{(0)}_q=(T_h-T_l)\left[S_{cl}(B)-S_{cl}(A)\right],
\end{equation}
\begin{equation}\label{eq:043}
\mathscr{W}^{(1)}_q=0,
\end{equation}
\begin{equation}\label{eq:044}
\mathscr{W}^{(2)}_q=(T_h-T_l)\left[S^{(2)}_{q}(B)-S^{(2)}_{q}(A)\right].
\end{equation}
Here, ${S_{cl}(A)}$ (${S_{cl}(B)}$) is the classical Gibbs entropy of the classical working substance before (after) the isothermal expansion process. One can see that ${\mathscr{W}^{(0)}_q}$ is nothing but the net work extraction during a classical Carnot cycle. ${S^{(2)}_{q}(A)}$ and ${S^{(2)}_{q}(B)}$ are the lowest order corrections of the entropy before and after the isothermal expansion process, which vanishes in the high temperature limit (${\beta\to 0}$) but are non-negligible at extremely low temperature. For some specific models, the quantum correction terms can be further simplified. For example, if we choose the quantum harmonic oscillator with an adjustable angular frequency ${\lambda(t)}$ as the working substance,
\begin{equation}\label{eq:045}
U(x,\lambda_t)=\frac{1}{2}m\lambda^2(t) x^2,
\end{equation}
one can get the lowest order quantum correction to the work extraction through Eq. (\ref{eq:044}),
\begin{equation}\label{eq:046}
\mathscr{W}^{(2)}_q=(T_h-T_l)\frac{\beta^2_h}{24}\left[\lambda^2(A)-\lambda^2(B)\right],
\end{equation}
which is consistent with the results in Ref. \cite{Quan2007} (one can see the consistency by expanding Eq. (56) in Ref. \cite{Quan2007} in powers of ${\hbar}$ to the second order). For other complicated potentials, such as a quartic oscillator \cite{Quan2015}, one can also get the lowest order quantum correction to the net work extraction through Eq. (\ref{eq:044}).
\section{Dicussion and Summary}
Before concluding this paper, we want to emphasize two points. First, it may be extremely hard to calculate the von Neumann entropy through its definition Eq. (\ref{eq:004}), because one have to diagonalize an infinite-dimensional matrix in order to compute the trace of a function. But reformulating the problem in the phase space significantly simplifies the calculation. Our results Eqs. (\ref{eq:020}-\ref{eq:022}) will be useful when studying quantum corrections to the entropy for complicated systems. Second, our results Eqs. (\ref{eq:020}-\ref{eq:022}) are not limited to thermodynamic equilibrium states, as we have clarified in Sec. III.

In summary, in this paper, we study the quantum corrections to the entropy in quantum systems. We expand the von Neumann entropy in powers of ${\hbar}$ by using the phase-space formulation. The zeroth order term reproduces the classical Gibbs entropy of its classical counterpart. Specially, for the thermodynamic equilibrium state, we verify our results of quantum corrections to the entropy, and we find that all correction terms odd in ${\hbar}$ are strictly zero. As an application, we derive the quantum corrections to the net work extraction during a quantum Carnot cycle.

In the classical stochastic thermodynamics, fluctuating entropy is defined along every stochastic trajectory in the phase space \cite{Seifert2005}. Nevertheless, how to define a trajectory-dependent entropy in quantum systems is still an open question. We plan to extend our current investigation to this regime and we believe that further studies along this line will advance our understanding about the relation between the quantum and the classical entropy and may bring important insights to some fundamental problems in quantum thermodynamics.

\section{Acknowledgment}
H. T. Quan acknowledges support from the National Science Foundation of China under grants 11775001, 11534002, and 11825001.


\begin{thebibliography}{99}
\bibitem{footnote2}
One can see that on the left-hand side of Eq. (\ref{eq:009}), there is no term in negative powers of ${\hbar}$ in both (${aW+b}$) and the Moyal product. And on the right-hand side of Eq. (\ref{eq:009}), there is no term in negative powers of ${\hbar}$ in ${W}$ either. Thus we can conclude that there is no term in negative powers of ${\hbar}$ in ${A_w}$.
\bibitem{footnote3}
From Eq. (\ref{eq:006}) and Eq. (\ref{eq:012}), one can see that there is no term in negative powers of ${\hbar}$ in ${f_w}$.
\bibitem{footnote4}
Some quantum systems do not have well-defined classical counterparts, e.g., the transverse Ising chain in an external magnetic field.
\end{thebibliography}
\end{document}